\documentclass[aps,twocolumn,showpacs]{revtex4}
\usepackage{amsfonts}
\usepackage{amsmath}
\usepackage{amssymb}
\usepackage{graphicx}
\usepackage{multirow}

\begin{document}


\title{The Density Matrix Renormalization Group and the Shell Model}

\author{S. Pittel and B. Thakur}

\affiliation{Bartol Research Institute and Department of Physics
and Astronomy, University of Delaware, Newark, Delaware 19716,
USA}

\date{\today}

\begin{abstract}
We summarize our recent efforts to develop the Density Matrix
Renormalization Group (DMRG) method into a practical truncation
strategy for large-scale nuclear shell model calculations.
Following an overview of the essential features of the DMRG, we
discuss the changes we have implemented for its use in nuclei. In
particular, we have found it useful to develop an angular-momentum
conserving variant of the method (the JDMRG). We then summarize
the principal results we have obtained to date, first reporting
test results for $^{48}$Cr and then more recent test results for
$^{56}$Ni. In both cases we consider nucleons limited to the 2p-1f
shell. While both calculations produce a high level of agreement
with the exact shell model results, the fraction of the complete
space required to achieve this high level of agreement is found to
go down rapidly as the size of the full space grows.

\end{abstract}

 \maketitle

\section{Introduction}

In the traditional nuclear shell model, the low-energy structure
of a nucleus is described by diagonalizing the effective nuclear
hamiltonian in an active space consisting of at most a few major
shells outside/inside an assumed doubly-magic core. Despite the
enormous truncation inherent in this approach, it can still only
be applied in very limited nuclear regimes. For heavy nuclei or
nuclei far from closed shells, further truncation of the
shell-model space to a manageable size is required.

In this work, we discuss the use of the Density Matrix
Renormalization Group (DMRG) as a truncation strategy for the
nuclear shell model. Originally developed for the treatment of
low-dimensional quantum lattices \cite{White}, the DMRG method was
subsequently extended with impressive success \cite{Duke1},
\cite{QC}, \cite{2D} to finite Fermi systems, suggesting its
possible usefulness for the nuclear shell model.

The DMRG method involves a systematic inclusion of the degrees of
freedom of the problem. When treating quantum lattices, real-space
sites are added iteratively. In finite Fermi systems, these sites
are replaced by single-particle levels. At each stage, the group
of sites that has been treated (referred to as a {\em block}) is
enlarged to include an additional site or level. This {\em
enlarged block} is then coupled to the rest of the system (the
{\em medium}) giving rise to the {\em superblock}. For a given
eigenstate of the superblock (often the ground state) or perhaps
for a group of important eigenstates, the reduced density matrix
of the enlarged block in the presence of the medium is constructed
and diagonalized and those states with the largest eigenvalues are
retained. This method of truncation is guaranteed to be optimal in
the sense that it maximizes the overlap of the truncated wave
function with the superblock wave function prior to truncation.

This process of systematically growing the system and determining
the optimal structure within that enlarged block is carried out
iteratively, by sweeping back and forth through the sites, at each
stage using the results from the previous sweep to define the
medium. In this way, the process iteratively updates the
information on each block until convergence from one sweep to the
next is achieved. Finally, the calculations are carried out as a
function of the number of states retained in each block, until the
changes are acceptably small.

The traditional DMRG method works in a product space, whereby the
enlarged block is obtained as a product of states in the block and
the added site and the superblock is obtained as a product of
states in the enlarged block and the medium. In the nuclear
context, this means working in the m-scheme. This method was
applied in nuclear physics by Papenbrock and Dean
\cite{Papenbrock} and shown to provide an accurate description of
the properties of $^{28}Si$, but a much poorer description of
$^{56}Ni$ where the converged solution was still energetically
quite far from the exact ground state.

A limitation of the traditional algorithm is that it does not
preserve symmetries throughout the iterative process. Since the
density matrix procedure involves a truncation at each iterative
stage, there is the potential to lose these symmetries and the
associated correlations. On this basis, we proposed \cite{Review}
the adoption of a strategy in which angular momentum is preserved
throughout the DMRG process.  This method, called the JDMRG, was
applied in nuclear physics for the first time in the context of
the Gamow Shell Model \cite{PloPlo} and then subsequently
developed as a means of approximating the traditional nuclear
shell model, with a preliminary application first reported
\cite{PS} for $^{48}Cr$. More recently, the algorithm was
computationally improved and applied both to $^{48}Cr$ and to the
ground state of $^{56}Ni$ \cite{TPS}. In this work, we extend the
results of ref. \cite{TPS} to include excited states of $^{56}Ni$.

An outline of the presentation is as follows. We begin in Section
II  with a brief overview of the traditional DMRG method including
a discussion of the changes required for exact angular-momentum
conservation. In Section III, we report our results for $^{48}Cr$,
which were first presented in \cite{TPS}, and then report our most
recent results for $^{56}Ni$, including those for excited states.
Finally, in Section IV we summarize the principal conclusions of
this work and outline some directions for future investigation.

\section{An overview of the DMRG method}

\subsection{The truncation strategy}

The DMRG method is based on an iterative inclusion of the degrees
of freedom of the problem, represented as sites on a lattice. This
is illustrated schematically in Figure 1 for a system with $8$
{\it ordered} sites.

\begin{figure}
\includegraphics[height=.2\textheight]{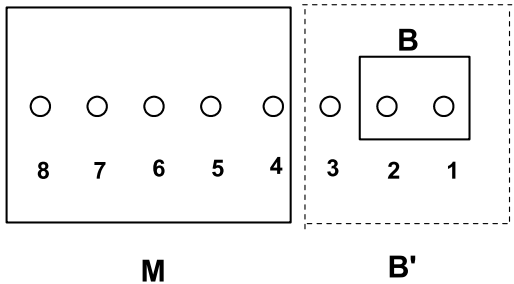}
\vspace{0.25cm} \caption{Schematic illustration of the DMRG growth
procedure. A
  block $B$ consisting of sites $1$ and $2$ is enlarged to include site
  $3$, forming $B'$. The medium $M$ consists of all of the remaining
  sites, $4~-~8$.
   }
\end{figure}

Assume that we have treated a group of sites, referred to as the
block and denoted $B$, and that we have retained a total number of
(optimal) states $m$ within that block.  We now wish to add to
this block the next site ($r$) with $l$ states, thereby producing
an enlarged block $B'$. For the moment, we will assume a product
(or m-scheme) description, so that the enlarged block has $m
\times l$ states,
\begin{equation}
|i,j>_{B'}=|i>_B |j>_r ~~~, ~~~i=1,m ~~, ~~j=1,l ~.
\end{equation}
We wish to retain the optimal $m$ states for the enlarged block.
How do we choose them?

In the DMRG method, we consider the enlarged block in the presence
of a medium $M$ that reflects all of the other sites of the
system, referring to the full system as the superblock (SB).
Assuming that the medium is likewise described by its optimum $m$
states, the $m \times l \times m$ states of the superblock can be
expressed as
\begin{equation}
|i,j,k>_{SB} = |i,j>_{B'} |k>_M ~.
\end{equation}

We then diagonalize the full hamiltonian of the system in the
superblock, isolating on its ground state,
\begin{equation}
|GS>_{SB}= \sum_{i,j,k} \Psi_{ijk} |i,j,k>_{SB} ~. \label{GS}
\end{equation}
If we then construct the reduced density matrix of the enlarged
block in the ground state,
\begin{equation}
\rho_{ij,i'j'} = \sum_k \Psi^*_{ijk} \Psi_{i'j'k}~,
\end{equation}
diagonalize it and retain the $m$ eigenstates with the largest
eigenvalues we are {\it guaranteed} to have the $m$ most important
(or optimal) states of the enlarged block in the ground state
(\ref{GS}) of the superblock.

It is straightforward to target a group of states of the system,
and not just the ground state, by constructing a mixed density
matrix containing information on all of them.

Once the optimal $m$ states are chosen, we renormalize all
required operators of the problem to the truncated space and store
this information. This includes all sub-operators of the
hamiltonian,
\begin{displaymath}
a^{\dagger}_i, ~ a^{\dagger}_ia_j, ~ a^{\dagger}_ia^{\dagger}_j,
~a^{\dagger}_ia^{\dagger}_ja_k, ~a^{\dagger}_ia^{\dagger}_ja_ka_l,
~+h.c.
\end{displaymath}
Having this information for the block and the additional level or
site enables us to calculate all such matrix elements for the
enlarged block as needed in the iterative growth procedure.

\subsection{Stages of the DMRG method}

With the above remarks as background, the DMRG procedure involves
the following stages.

\subsubsection{The warmup stage}
In the warmup stage we make an initial guess on the optimal $m$
states for each block. This choice is important in determining how
rapidly the iterative procedure converges. In our treatment, we do
this by growing blocks from each side of the chain gradually,
using those orbits that have already been treated on the other end
as the medium. This is illustrated schematically in Figure 2 for
two successive steps of the warmup procedure.

\begin{figure}
\includegraphics[height=.35\textheight]{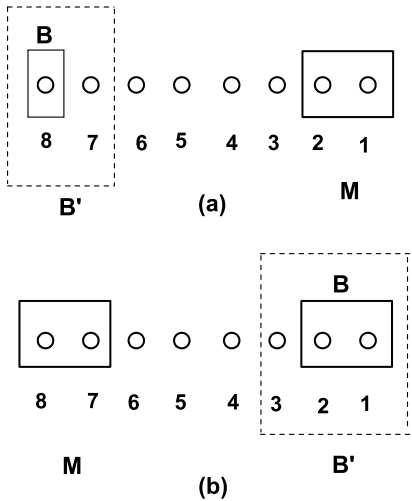}
\vspace{0.25cm} \caption{Schematic illustration of two successive
steps in the warmup procedure. (a)  A block involving the site $8$
is enlarged to include the next site $7$, with the medium being a
block involving the optimum states from the sites $1$ and $2$
obtained in the previous warmup step. (b) A block involving the
sites $1$ and $2$ is enlarged to include the next site $3$, with
the medium being a block involving the optimum states from the
sites $8$ and $7$ obtained in the warmup step of part (a).
   }
\end{figure}

\subsubsection{The Sweep stage}

\begin{figure}
\includegraphics[height=.35\textheight]{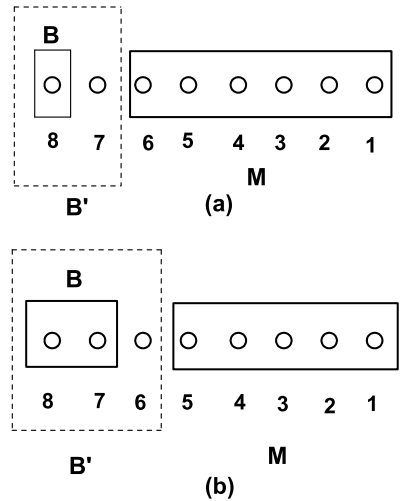}
\vspace{0.25cm} \caption{Schematic illustration of two successive
steps in the sweep procedure. (a)  A block involving the site $8$
is enlarged to include the next site $7$, with the medium being a
block involving the optimum states from the sites $1-6$ obtained
previously. (b) A block involving the sites $8$ and $7$ is
enlarged to include the next site $6$, with the medium being a
block involving the optimum states from the sites $1-5$ obtained
previously.
   }
\end{figure}

In this stage of the process, schematically illustrated in Figure
3, we gradually sweep back and forth through the sites of the
lattice, at each step of the sweep using for the medium the
results either from the warmup phase (during the first sweep) or
from the previous sweep stage. The sweep process is done over and
over until convergence is achieved from one sweep to the next.

\subsubsection{As a function of $m$}

The above calculations are done for a given choice of $m$. The
calculations are then done as a function of $m$, until the changes
with increasing $m$ are acceptably small.

\subsection{The JDMRG approach}

As noted earlier, most DMRG approaches violate symmetries. In
nuclei, for example, they typically work in the m-scheme. Such a
procedure is potentially problematic when imposing truncation,
however, as it is difficult to ensure that the states retained
contain all the components required by the Clebsch Gordan series
to build states of good angular momentum. For this reason, we have
chosen to develop an angular-momentum-conserving variant of the
DMRG method in which angular momentum is preserved throughout the
growth, truncation and renormalization stages. The most
significant difference between this (the JDMRG) approach and the
more traditional DMRG approach is that now we must calculate and
store the {\em reduced matrix elements} of all sub-operators of
the hamiltonian,

\begin{eqnarray*}
&& a^{\dagger}_i, ~ [a^{\dagger}_i\tilde{a}_j]^K, ~
[a^{\dagger}_ia^{\dagger}_j]^K, ~\left( [a^{\dagger}_ia^{\dagger}_j]^K\tilde{a}_k\right)^L, \nonumber \\
&& ~\left([a^{\dagger}_ia^{\dagger}_j]^K
~[\tilde{a}_k\tilde{a}_l]^K \right)^0~ +h.c.
\end{eqnarray*}
This can be done using standard Racah techniques.

\subsection{A three-block JDMRG strategy}

In the nuclear shell-model calculations we will report here, we
adopt a three-block strategy for the enlargement and truncation
process, schematically summarized in Figure 4.

\begin{figure}[htb]
\includegraphics[height=.15\textheight]{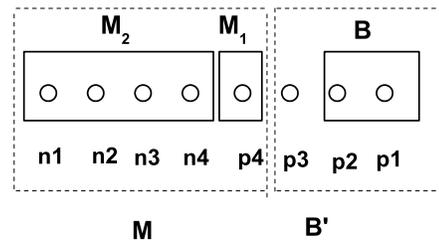}
\vspace{0.25cm} \caption{Schematic illustration of the three-block
DMRG growth procedure for a system with neutron and proton levels.
 }
\end{figure}

We begin by choosing our order of sites so that neutron and proton
orbitals sit on opposite ends of the chain. We then gradually grow
blocks of each type of particle only, namely we grow neutron
blocks and proton blocks but no mixed blocks. Lastly, in the sweep
stage we go back and forth through the orbits of a given type of
particle only. As can be seen from the figure, the medium in this
approach involves two components. If, for example, we are
enlarging a proton block, the full medium ($M$) involves all of
the remaining proton levels ($M_1$) and all of the neutron levels
($M_2$).

\section{Calculations}

We have carried out test calculations of the JDMRG method
described above on the nuclei $^{48}Cr$ and $^{56}Ni$. We assume
that these nuclei can be described in terms of valence neutrons
and valence protons outside a doubly-magic $^{40}Ca$ core. In
both, we use the order of single-particle levels shown in Figure
5.

We report the results for these two applications in the following
subsections.

\subsection{Results for $^{48}Cr$}

We begin with our results for $^{48}Cr$, for which there are four
neutrons and four protons outside a $^{40}Ca$ core. In these
calculations we assumed a KB3 effective interaction between
valence nucleons and compared our results with those obtained for
the same hamiltonian in ref. \cite{Poves1}.  The size of the full
shell-model space in this case involves 1,963,461 states, of which
41,355 are $0^+$ states, 182,421 are $2^+$ states, 246,979 are
$4^+$ states, etc.

Our results for the ground state are presented in Table I. The
exact calculation produces a ground state energy of $-32.953~MeV$.
The DMRG calculation converges smoothly to this result as $m$ is
increased, but requires the inclusion of a substantial fraction of
the full space to obtain a high level of accuracy. With of order
25\% of the full $0^+$ space, we are able to achieve accuracy to
only a few $keV$. To achieve an accuracy of better than 50 $keV$
still requires, however,  over 20 \% of the full $0^+$ space.

In Figure 5, we show how the results converge as a function of the
number of sweeps.  As can be readily seen, after the first sweep
we are extremely close to the final converged result. This is a
direct consequence of our use of a warmup procedure that
incorporates (in step $0$) a significant part of the correlations.

\begin{figure}[htb]
\vspace{0.2in}
\includegraphics[height=.15\textheight]{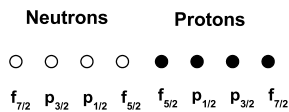}
 \caption{Order of single-particle levels used in
the DMRG growth algorithm for both $^{48}Cr$ and $^{56}Ni$.
 }

\end{figure}

\begin{table}[htb]
\caption{Calculated ground-state energies in $MeV$ as a function
of $m$ for $^{48}$Cr. The maximum dimension encountered in the
sweep process is also given. }
\begin{tabular}{|c|c|c|}
\hline
  $m$  &$ E_{GS}$ &   $Max~Dim$\\
\hline
40  & -32.698  &  ~1,985 \\
60  & -32.763  &  ~2,859 \\
80  & -32.788  &  ~3,765 \\
100 & -32.817  &  ~4,494  \\
120 & -32.840  &  ~6,367  \\
140 & -32.890  &  ~8,217  \\
160 & -32.902  &  ~9,978  \\
180 & -32.944  & 11,062  \\
200 & -32.947  & 12,076  \\
\hline
Exact & -32.953 & 41,355 \\
\hline
\end{tabular}
\end{table}

In Table II, we present results for the lowest excited states,
obtained using the blocks obtained at the ground state minimum.
Here too convergence to the exact results is achieved, if a
sufficiently large fraction of the full shell-model space is
retained. This is true despite the fact that the density matrix
truncation procedure implemented targeted the ground state only.

\begin{table}
\caption{Results for the excitation energies in $MeV$ of the
lowest $2^+$ and $4^+$ states in $^{48}Cr$ from the JDMRG
calculations described in the text. The dimensions of the
associated hamiltonian matrices are given in parentheses. }
\label{tab:a}
\begin{tabular}{|c|c|c|}
\hline
$m$ &$2_1^+$ & $4_1^+$ \\

 \hline
140 & 0.873 (~~9,191) & 2.022 (~12,442) \\
160 & 0.860 (~12,038) & 1.996 (~16,553) \\
180 & 0.855 (~15,148) & 1.989 (~21,628) \\
\hline
$Exact$ & 0.806 (182,421) & 1.823 (246,979) \\
\hline
\end{tabular}
\end{table}

\subsection{$^{56}Ni$}

Next we turn to $^{56}Ni$, for which the size of the full
shell-model space in the m-scheme contains 1,087,455,228 states,
of which  15,443,685 have $J^{\pi}=0^+$, 71,109,189 have
$J^{\pi}=2^+$ and 105,537,723 have $J^{\pi}=4^+$, in all cases
much larger than in $^{48}Cr$.

In our earlier published work on $^{56}Ni$, we used the KB3
interaction for $^{56}Ni$ as well and compared the ground state
energy with the results from ref. \cite{Poves2}. We only
considered the ground state in those calculations, since those
were the only exact results that had been reported in the
literature. The reason for that is that KB3 is known to produce
poor agreement with the experimental spectra for nuclei in this
region of the $2p-1f$ shell. Much better agreement can be obtained
with the improved GXPF1A interaction \cite{GXPF1A}. Thus, we have
redone our test calculations of $^{56}Ni$ with this interaction,
now comparing with the exact results not only for the ground state
but for low-lying excited states as well \cite{Bro06}. These are
the largest test calculations we have carried out to date using
the JDMRG method.

\begin{figure}[htb]
\includegraphics[height=.25\textheight]{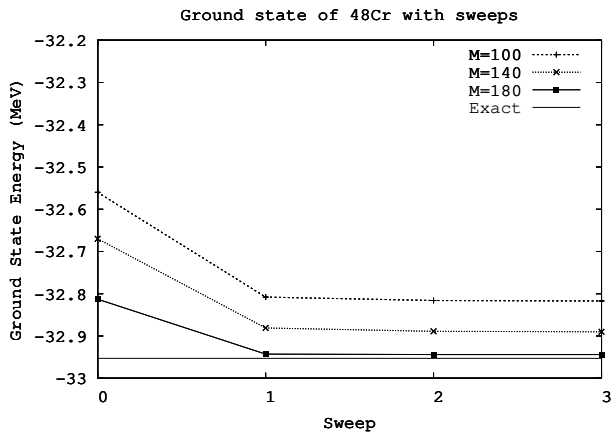}
\caption{Convergence of the results for $^{48}Cr$ with the number
of sweeps.
   }
\end{figure}

The results for the ground state energy as a function of $m$ are
shown in Table III. Here we are able to achieve roughly 60 $keV$
accuracy with just 0.5\% of the full $0^+$ space. This is a
significantly lower fraction of the full space than was required
in $^{48}Cr$ to achieve the same level of accuracy.

\begin{table}[b]
\caption{Results for the ground-state energy of $^{56}Ni$ in $MeV$
from the JDMRG calculations described in the text. {\em Max~Dim}
refers to the maximum dimension of the superblock hamiltonian
matrix. } \label{tab:c} \vspace{0.1cm}
\begin{tabular}{|c|c |c|}
\hline
$m$ &$E_{GS}$ & $Max~Dim$ \\

 \hline
60 & -205.603 & ~~~~64,397 \\
80 & -205.632 & ~~~~74,677 \\
100 & -205.643 & ~~~~87,633 \\
120 & -205.652 & ~~~~106,383 \\
\hline
$Exact$ & -205.709 & 15,443,685 \\
\hline
\end{tabular}
\end{table}

This is perhaps the most encouraging result of our work to date.
It suggests that the fraction of the space required to achieve a
high level of accuracy with the JDMRG method goes down rapidly as
the size of the space problem increases. If this is confirmed with
extension to even larger problems it would bode very well for the
future usefulness of the JDMRG method as a practical truncation
approach for large-scale shell-model studies.

As noted above, exact results also exist with the GXPF1A
interaction for excited states. In Table IV, we present our
results for the lowest $2^+$ and $4^+$ states, again in comparison
with the exact results. These results were obtained by
diagonalizing the associated hamiltonian matrices that derive with
the blocks obtained at the ground state minimum. Here too the
results are getting better with $m$, albeit slowly. Agreement with
the exact results for $m=100$ is still not as good as we would
like. It might be possible to improve the quality of the
description of low-lying excited states by targeting both the
ground state and the first excited state in the density matrix
truncation procedure, hopefully without losing too much accuracy
in our reproduction of the ground state.

\begin{table}
\caption{Results for the excitation energies in $MeV$ of the
lowest $2^+$ and $4^+$ states in $^{56}Ni$ from the JDMRG
calculations described in the text. The dimensions of the
associated hamiltonian matrices are shown in parentheses. }
\label{tab:d} \vspace{0.1cm}
\begin{tabular}{|c|c|c|}
\hline
$m$ &$2_1^+$ & $4_1^+$ \\

 \hline
60 & 2.970 (~~~296,633) & 4.137 (~~~~445,898) \\
80 & 2.944 (~~~345,213) & 4.123 (~~~~556,572) \\
100 & 2.942 (~~~423,265) & 4.090 (~~~~701,502) \\
\hline
$Exact$ & 2.600 (71,109,189) & 3.688 (105,537,723) \\
\hline
\end{tabular}
\end{table}

\section{Summary and Outlook}

In this talk, we have summarized the current status of our efforts
to build the Density Matrix Renormalization Group Method method
into a practical truncation strategy for large-scale shell-model
calculations of atomic nuclei. Following an overview of the
essential features of the method,  we discussed the changes we had
to implement for its application to nuclei. Most importantly, we
found it useful to develop an angular-momentum conserving version
of the method, the JDMRG. We then summarized the results we have
obtained for the nuclei $^{48}$Cr and $^{56}$Ni, in both cases
comparing with the results of exact diagonalization. Both
calculations were able to accurately reproduce the exact
shell-model results. In the case of $^{48}Cr$, however, this high
level of accuracy required us to retain a very large fraction of
the full space.  In contrast, we were able to achieve comparably
accurate results for $^{56}Ni$ with a much smaller fraction of the
space. The fact that the fraction of the space goes down rapidly
with the size of the problem bodes well for the future usefulness
of the method in even larger shell-model problems.

There are several issues that we intend to explore in the near
future. One concerns the need to determine through additional
calculations how rapidly the fraction of the space required for
convergence scales with the size of the problem. Currently we only
have two data points, $^{48}Cr$ and $^{56}Ni$. More are needed to
draw meaningful conclusions.

We are also looking into the question of whether we can obtain
comparable agreement for both the ground state and low-lying
excited states by building a mixed density matrix that includes
information not only on the ground state but also on the first
excited $0^+$ state.

We are also in the process of applying these methods to odd-mass
nuclei around $^{56}Ni$.  Exact and Coupled Cluster results are
now available for several such nuclei \cite{Horoi} and we are
interested in seeing how well the JDMRG truncation strategy works
on these nuclei and how well it compares with the Coupled Cluster
method.

Another issue of current interest concerns the possibility of
breaking up large single-particle orbitals, rather than adding
them in a single stage. We have some thoughts on how this might be
done without losing angular momentum conservation and we are now
testing these ideas on the $f_{7/2}$ orbital for the same nuclei
we have already studied.  If successful, we will then consider the
application of these ideas to even larger shells, as will be
critical for subsequent applications of the method to heavier
nuclei, the ultimate goal of this project.

\vspace{0.2in}

\begin{flushleft}
{\bf Acknowledgments}
\end{flushleft}
This work is based on a talk presented by one of the authors
(S.P.) at the XXXIInd International Symposium on Nuclear Physics
held in Cocoyoc, Mexico from 5-9 January 2009. It was supported by
the US National Science Foundation under grant \# PHY-0553127. We
thank Jorge Dukelsky for his significant involvement in the
development of the JDMRG method, Nicu Sandulescu for his major
contributions to much of the work reported here, and Alfredo Poves
for providing us with the KB3 matrix elements and $^{48}Cr$ exact
results we reported.

\end{document}